\documentclass{aa} 
\usepackage{natbib}

\usepackage{graphicx}
\usepackage{txfonts}
\begin{document}
   \title{Reddening and metallicity maps of the Milky Way bulge from VVV and 2MASS\thanks{Based on observations taken within the ESO VISTA Public Survey VVV, Programme ID 179.B-2002}}
   \subtitle{III. The first global photometric metallicity map of the Galactic bulge}
   \author{O. A. Gonzalez$^{1,4}$  \and M. Rejkuba$^{2}$ \and M. Zoccali$^{3,4}$ \and E. Valenti$^{2}$ \and D. Minniti$^{3,4,5}$ \and R. Tobar$^{2}$}
   
   \offprints{O. A. Gonzalez}

   \institute{ $^{1}$European
   Southern Observatory, Casilla 19001, Santiago 19, Chile\\ \email{ogonzale@eso.org}\\
$^{2}$European Southern Observatory, Karl-Schwarzschild-Strasse 2, D-85748 Garching, Germany\\ \email{mrejkuba@eso.org; evalenti@eso.org}\\
$^{3}$Departamento    Astronom\'ia    y Astrof\'isica, Pontificia Universidad  Cat\'olica de Chile,  Av. Vicu\~na Mackenna
   4860,         Stgo.,         Chile\\         \email{mzoccali@astro.puc.cl;
dante@astro.puc.cl}\\
$^{4}$The Milky Way Milennium Nucleus, Av. Vicu\~na Mackenna 4860, 782-0436 Macul, Santiago, Chile\\
$^{5}$Vatican Observatory, V00120 Vatican City State, Italy\\
}
   \date{Received / Accepted}

  
\abstract  
{}
{We investigate the large scale metallicity distribution in the Galactic bulge, using a large spatial coverage, in order to constrain the bulge formation scenario.}
{We use the VISTA variables in the Via Lactea (VVV) survey data and 2MASS photometry, covering 320 sqdeg of the Galactic bulge, to derive photometric metallicities by interpolating of the $(J-Ks)_0$ colors of individual Red Giant Branch stars based on a set of globular cluster ridge lines. We then use this information to construct the first global metallicity map of the bulge with a resolution of $30'\times45'$.}
{The metallicity map of the bulge revealed a clear vertical metallicity gradient of $\sim0.04 $dex/deg ($\sim0.28 $dex/kpc), with metal-rich stars ([Fe/H]$\sim$0) dominating the inner bulge in regions closer to the galactic plane ($|b|<5$). At larger scale heights, the mean metallicity of the bulge population becomes significantly more metal-poor.}
{This fits in the scenario of a boxy-bulge originated from the vertical inestability of the Galactic bar, formed early via secular evolution of a two component stellar disk. Older, metal-poor stars dominate at higher scale heights due to the non-mixed orbits from the originally hotter thick disk stars.}
             
\authorrunning{Gonzalez et al.}
\titlerunning{The complete photometric metallicity map}

\maketitle

%
\section{Introduction}

In the hierarchical scenario of galaxy formation, the early mass assembly of galaxies will be dominated by merger processes which favors the formation of spheroids \citep{toomre77}. This violent and relatively fast process produces the present day elliptical galaxies. However, when the spheroids are accompanied by the (re-)growth of stellar disk \citep{kauffmann99, springel05}, they receive the name of classical bulges. 

Classical bulges are expected to share the same observational properties of elliptical galaxies. Their light profiles are well fitted by a Vaucouleurs-like profile with n$\sim$4 \citep[e.g.][]{gadotti09} and their stars show hot dynamics dominated by velocity dispersion \citep[e.g.][]{gadotti11}. An important point is that the rotation of classical bulges is observed to be non-cylindrical, therefore showing a decrease in the rotation velocity at increasing height from the plane \citep[e.g.][]{emsellem04}. Since classical bulges are formed violently, in fast and early processes of collisions and mergers, stars in classical bulges are expected to have formed mostly from the starburst in the early times, with little or no star formation since then. For this reason, classical bulges are expected to be composed predominantly of old stars ($\sim$10 Gyr) \citep[e.g.][]{renzini98}, they have enhanced abundance ratios \citep{Peletier99, moorthy06, macarthur08} and often show metallicity gradients, going from metal-poor stars in the outskirts to a more metal-rich population towards the center. \citep[see for example][]{jablonka07}.

Given that the classical bulge formation scenario is connected to the rate of merger events, this is expected to be the dominant process in the evolution of galaxies at early times \citep{kormendy04}. Later on, secular evolution processes would become important, re-arranging stellar material from a settled galactic disk. Disks instabilities, such as the presence of spiral arms, are expected to induce the formation of bars in the inner regions of a galactic disk \citep{combes81,weinberg85,debattista98, athana02}. Indeed, galactic bars are a very common phenomena, found in $\sim 60\%$ of spiral galaxies in the local Universe \citep[e.g.][]{eskridge00}. 

Although disks are thin galactic components, bars might become thicker because of the vertical resonances and buckle off the plane of the galaxy. For this reason, galaxies,  when observed edge-on, often show a central component which swells out of the disk with a boxy, peanut, or even an X-shape \citep[e.g.][]{bureau06,patsis02}. These structures have been seen in numerous simulations, as the result of the development of vertical instabilities from the, initially thin, stellar bar \citep{athana05, debattista04, debattista06, martinez06}.

The dynamical and structural properties of boxy-bulges are well constrained from models and observations. However, the properties of their stellar populations are less clear than in the case of classical bulges. In fact, a variety of stellar populations are observed within bars \citep[e.g.][]{gadotti06, perez09}. In practice, since boxy-bulges are bars thickened by vertical instabilities, the null hypothesis would be that they should hold similar ages as their parent disks. Similar case would be expected for the metallicity and alpha elements abundances. However, \citet{freeman08} suggested that age and abundance gradients could be present in boxy bulges. Older, more metal-poor stars would have more time to be scattered to larger distances from the plane by the buckling process, therefore establishing a negative metallicity and positive age gradient. The presence and strength of such gradients would then depend on the chemical enrichment history of the parent disk material, and the timescale of the buckling instability to form the boxy-bulge.

The brightness profiles of boxy-bulges are less centrally concentrated than those of classical bulges, and are better fitted by light profiles with a lower sersic index \citep[e.g.][]{kormendy78,gadotti09}. What is probably the most characteristic property of boxy-bulges its that they present cylindrical rotation patterns. This means that the rotation velocity within a boxy-bulge does not change with height above the galactic plane \citep[e.g.][]{kormendy82, combes90, athana02, shen10}.

The presence of a central bar has important consequences for the subsequent evolution of the galaxy, redistributing the angular momentum and matter in the disk. In particular, it causes the loss of angular momentum of the molecular gas settled in the rotating disk, inducing its infall process into the inner regions of the galaxy. The gas, slowly fueling into the center, produces a slow star formation process that results in the formation of a disk-like component in these inner regions. We refer to these structures as pseudo-bulges. The brightness of pseudo-bulges can be fitted with a very disk-like, exponential light profile and they are observed as flat, rotation supported structures. Also, as a result of a very slow process of gas fueling into the center because of the presence of the bar, the star formation is expected to be slow and continuous. A direct consequence is that pseudo-bulges host young stellar populations and are dominated by metal-rich stars with solar alpha element abundances.

As described before, boxy-bulges are thickened bars, with bars being in fact disk phenomena resulting from internal processes of secular evolution. The absence of an external factor in the formation process of boxy-bulges and the fact that they are both formed our of originally disk material, as opposite to the classical bulge scenario, led different authors to refer to them as pseudo-bulges. This was also qualitatively supported by the pseudo-bulge definition in the review of \citet{kormendy04}, where they describe:\textit{'..if the component in question is very E-like we call it bulge, if its more disk-like, we call it pseudo-bulge'}. The morphology of thickened bars is certainly not E-like \citep{bureau06}. However, to avoid confusion during our analysis, we refer to them as boxy-bulges and not pseudo-bulges.

The above scenarios for bulge formation and their properties, mostly based on the study of bulges in nearby galaxies, has been recently challenged by the observation of galaxies at redshift $\sim2$, i.e., at a lookback time comparable to the ages seen in classical bulges. At this redshift, disk galaxies are observed to hold large, rotating disks with a very high gas fraction and velocity dispersion, significatly different from local spirals \citep{genzel06,genzel08, fs09, tacconi10, daddi10}. Massive gas clumps are forming in these gas-rich galaxies via disk intestabilities, and then interact and likely merge in the center of the disk. This violent process naturally leads to the rapid formation of a  bulge over timescales of a few $10^8$ yr, much shorter than that typically
ascribed to secular instabilities in local disk galaxies \citep{immeli04, carollo07, elmegreen08}, and therefore results in an old, $\alpha$-element enhanced, present day stellar populations \citep{renzini09,peng10} of both bulge and disk formed at the early cosmic epoch.

With the exception of the Milky Way and M31 \citep{jablonka05}, the observational studies of the central components of galaxies, based on which most of the above formation scenarios are derived, are strongly challenged by the impossibility to resolve individual stars due to crowding and blending in high stellar density regions of distant galaxies. In this context, the Milky Way bulge, one of the major components of the Galaxy, can be studied at a unique level of detail thanks to its resolved stellar populations which hold the imprints of how our Galaxy formed and evolved. Thus, a better understanding of the Milky Way bulge will provide solid constraints to interpret observations of extragalactic bulges as well as for detailed galaxy formation models. 

The place of the Milky Way within these different bulge formation scenarios is far from being understood. The Galactic bulge structure is observed to be dominated by a stellar bar \citep{stanek97} oriented $\sim 25$ deg with respect to the Sun-Galactic center line of sight \citep{stanek97, lopez05, rattenbury07, gonzalez12, nataf12}. The observed age of the bulge stars is 10 Gyr or older, based on the deep photometric studies in few low extinction windows \citep{zoccali03, clarkson11}, but \citet{bensby12} found a non-negligible fraction of younger stars based on a microlensing study. A vertical metallicity gradient has been observed along the bulge minor axis \citep{zoccali08, johnson11} that most likely flattens in the inner regions \citep{rich_origlia07}. In addition to the population of stars associated with a metal-rich, alpha-poor stellar bar, there are hints for a distinct component dominated by more metal-poor, alpha enriched stars showing hot kinematics which resemble those of a classical bulge \citep{babusiaux10, hill11, gonzalez11_alphas, uttenthaler12}. Furthermore, significant evidence is now available suggesting that the Galactic bulge is X-shaped in the outer regions \citep{mcwilliam10, saito11}, as expected from buckling instability processes of the bar, although the metal-poor stars do not seem to follow the same spacial distribution \citep{ness12}.

Such complex set of properties for the Galactic bulge are the result of the detailed study of discrete fields, mostly concentrated along the minor axis, especially for the case of the chemical abundances. A larger, more general mapping of these properties is the necessary bridge between the detailed studies of the Galactic bulge and those of external bulges, where integrated light studies can only provide us with the global information.

In \citet{gonzalez11_I} (Paper I) we have shown a procedure to build multiband near-IR catalogs using the VISTA Variables in the Via Lactea (VVV) public survey data. We have then used these to derive a complete, high-resolution extinction map of the Galactic bulge, based on the color information of the red clump (RC) giant stars. The extinction map was presented in \citet{gonzalez12} (Paper II) together with a discussion of the important implications of its use in bulge studies. One of the improvements is the availability of a de-reddened RC feature in the luminosity function which allows us to trace the mean distance of stars towards different regions of the bulge, the so-called red clump method. The mean distance to each line of sight allows us then to place our photometric catalogs on the absolute plane and to derive photometric metallicity distributions. In Paper I we have tested the method by comparing these photometric metallicity distributions for regions along the bulge minor axis, where accurate spectroscopic metallicities are available, finding a remarkable agreement between both methods.  In this article, the last of the series, we extend the derivation of photometric metallicities to the rest of the region covered by the VVV survey. We provide for the first time the global view of the metal content of bulge stars and its gradients as if we were looking at the Milky Way as an external galaxy.    


\section{The data}

We use the VVV public DR1 catalogues  \citet{saito12}. For a complete description of the survey and the observation strategy we refer the reader to \citet{vvv10}. A detailed description of the combination of the tile-based source lists and their calibration is given in Paper I. Methods used in Paper I and Paper II are based on the analysis of the RC properties, which are restricted to a magnitude range well covered by the VVV survey photometry, across $\sim320$ sqdeg of the Galactic bulge, spanning from $-10 < l < 10$ and $-10 < b < +5$. For this final part of the project, we require upper RGB photometry to derive photometric metallicities by color interpolation between globular clusters ridge lines. As described in other articles \citep{gonzalez11_I, saito12}, VVV photometry suffers from saturation for stars brighter than $K_s=12$ which corresponds to the upper RGB region in the bulge CMDs required for the color interpolation. For this reason, in Paper I, we have constructed VVV multiband catalogs and calibrated them into the 2MASS system, in order to later replace the saturated range of magnitudes with those of 2MASS photometry. 

The VVV-2MASS calibration procedure is based on the comparison of stars flagged in 2MASS with high photometric quality in all three bands, in the magnitude range between $12<K_s<13$. This magnitude range avoids the saturated stars in VVV while being bright enough to ensure a sufficient number of 2MASS sources with accurate photometry for the calibration. The zero-point was calculated and applied individually to each VVV tile catalog. Finally stars with $K_s<12$ mag were removed from the VVV catalogue and were replaced with 2MASS sources. 

Since 2MASS is severly affected by crowding in the regions located close to the galactic plane, we restrict our analysis for latitudes $|b|>2$. This restriction aids also to avoid the high extinction regions where some differential reddening, in variation scales smaller than 2', the resolution of the bulge extinction map from Paper II.

\section{Extinction and mean distances}

Applying the extinction correction derived for the whole bulge on a spatial resolution of 2' $(|b|<4)$, 4' $(4<|b|<6)$, or 6' $(b>6')$ (Paper II) and using the extinction law of \citet{cardelli89}, the de-reddened luminosity function is derived for each line of sight in the bulge. The distances to these lines of sight are then derived by applying the equation:

\begin{equation}
(m-M)_0 =K_s(RC) - M_{K_s}(RC) - A_{K_s} + \Delta M_{K_s}
\end{equation}

$M_{K_s}=-1.55$ is adopted for the RC luminosity for 10 Gyr old population with a solar metallicity \citep{salaris+girardi02, pietri04, vanhel07}, which is an average value appropriate for the bulk of the Milky Way bulge stars \citep{zoccali03}. However, the intrinsic magnitude of the RC ($M_{K_s}$) is known to have a certain dependence on metallicity. Although this dependence is neglected by most studies, a large metallicity gradient present in the bulge could bias our own metallicity calculation if a single $M_{K_s}$ value is adopted for the whole bulge. In order to take into the account the (expected) presence of the metallicity variations we apply the metallicity correction term ($\Delta M_{K_s}$).  As a consequence, an iterative process, described in the next section, is adopted to derive the fully consistent distance and metallicity for each line of sight.

\section{Photometric metallicities}

The method to derive photometric metallicities in this work is based on the interpolation of the dereddened $(J-K_s)_0$ colors of upper RGB stars between ridge lines of globular clusters with well-know metallicities derived from spectroscopy. In Paper I we have compared the photometric metallicity distribution along the bulge minor axis, to that of high resolution spectroscopy for the same regions. We observed a very good agreement between both results, and in particular, the photometric results reproduced the minor axis metallicity gradient that has been confirmed in numerous specotroscopic studies. 

\begin{figure}[htb]
\begin{center}
\includegraphics[scale=0.70]{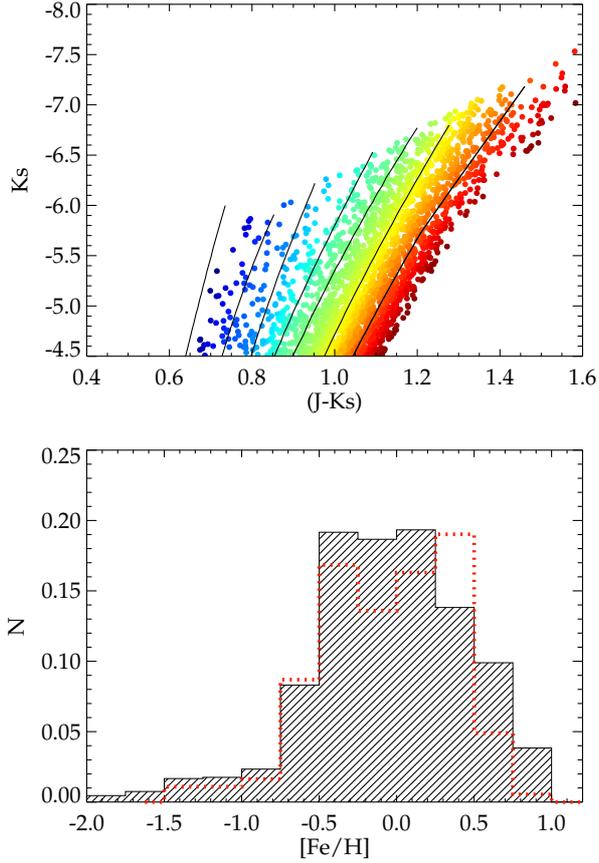}
\caption{Color-magnitude diagram of the Galactic bulge in Baade's window $(30'\times30')$, in the selected upper RGB region, used for the interpolation routine. Stars are color coded acording to their individual metallicities. Lower panel shows the MDF for the same region, compared to that from high-resolution spectroscopy from Hill et al. (2011) shown in red.}
\label{mdf_full}
\end{center}
\end{figure}

The complete method can be summarized as follows: 

\begin{enumerate}
\item The $K_{s_0}$ luminosity functions are built for each VVV tile and the mean magnitude of the RC is calculated by fitting the equation:

\begin{equation}
N(K_{s_0})=a+bK_{s_0}+cK_{s_0}^2+\frac{N_{RC}} {\sigma_{RC}\sqrt{2\pi}}\exp\Big[\frac{(K_{s_0}^{RC}-K_{s_0})^2}{2\sigma^2_{RC}}\Big]
\end{equation}

\item Each tile is divided in four subfields of $30'\times45'$. The distance modulus is then derived for each subfield by $(K_{s,0}-M_{K_s})_{RC}$ using equation 1. In the first step, $\Delta M_{K_s}=0$, but this value changes in subsequent iterations. 
\item The $(J-K_s)_0$ vs. $M_{K_s}$ CMD for each $30'\times45'$ sub-field is used to select stars with $M_{K_s}>-4.5$ and $(J-K_s)_0>0.65$, selecting the RGB region above the AGB bump that is more sensitive to metallicity, and avoiding most of the foreground disk contamination. An upper magnitude cut is also applied at the $M_{K_s}$ corresponding to the RGB tip measurements of bulge globular clusters from \citet[][]{valenti07}.
\item The color of each star is then interpolated among the empirical templates of Galactic halo and bulge globular clusters (M92, M55 , NGC 6752, NGC 362, M 69, NGC 6440, NGC 6528 and NGC 6791), with well known metallicity, selected from the sample of \citet[][]{valenti07}. An individual metallicity value is obtained for the stars in each subfield by this color interpolation.
\item The mean metallicity of the subfield is then obtained and the $\Delta M_{K_s}$ value is calculated,  using the new mean metallicity, based on \citet[][]{salaris+girardi02} equation:
\begin{equation}
\Delta M_{K_s}=0.275[Fe/H]+1.496-M_{K_s}(RC),
\end{equation}
and applied to equation 1.
\end{enumerate}

Steps 2-5 are iterated to obtain a final mean metallicity, that corresponds to a single $M_{K_s}$.
An example of the CMD in the Baade's window, color coded by the metallicity of the individual stars, is shown in the upper panel of Figure 1. The lower panel compares the photometric metallicities (hashed histogram) with the high resolution spectroscopic measurements from \citet{hill11} presented in red.

Having proven the reliability of the mean metallicity measurements from the VVV+2MASS photometry (Paper I) we can now present in Fig.~\ref{mdf_full}, the first global map of mean metallicity in bins of $30'\times45'$. This map is restricted to $|b|>3^\circ$ as the upper RGB in our dataset corresponds to 2MASS photometry which is limited by crowding in the inner regions. This is the first global view of the mean metallicity variations across the Milky Way bulge. The gradient along the minor axis, observed from high-resolution studies, is clearly seen in our map and it is the result of a central concentration of metal-rich stars in the region $-5^\circ<b<5^\circ$ and $-7^{\circ}<l<7^{\circ}$. This concentration of near-solar metallicity stars seems to follow a radial gradient of $\sim0.04$ dex/deg, going from more metal-poor stars in the outer bulge to more metal-rich stars towards the inner regions.

\begin{figure}[htb]
\begin{center}
\includegraphics[scale=0.50]{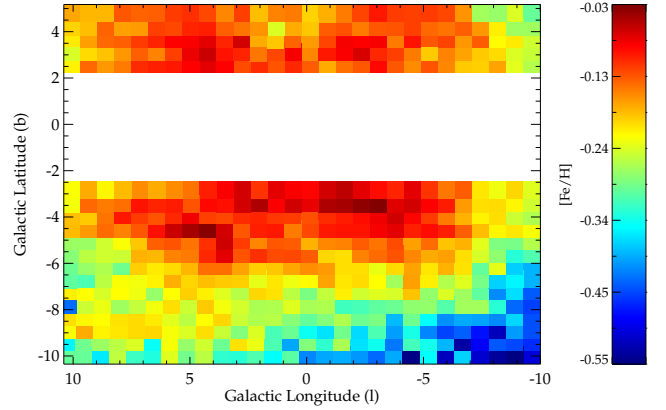}
\caption{Map of the mean values of the metallicity distributions for the Galactic bulge covered by the VVV survey using the Cardelli extinction law.}
\label{mdf_full}
\end{center}
\end{figure}

As described in Paper I, the metallicity map will be publicly available to the community via the BEAM (Bulge Extinction And Metallicity) calculator webpage\footnote{http://mill.astro.puc.cl/BEAM/calculator.php}.

\section{Sources of error}

Clearly, the most accurate abundance measurements are obtained using high-resolution spectroscopic methods. Unfortunately, such studies are very expensive in terms of telescope time, allowing only a few regions to be studied and thus not allowing us to obtain the large scale picture of the bulge chemical abundance properties. The accuracy of the photometric abundance derivations depend strongly on the knowledge of other properties affecting the stellar population under study such as the density distribution in the corresponding line of sight and the interstellar extinction. Although we have significantly improved our handling on the bulge general properties such as distance spread and interstellar extinction, possible variations across different regions might still be present. Therefore, it is important to quantify the effect of such variations on our mean abundance determinations.

\subsection{The bulge distance spread}

\begin{figure}[htb]
\begin{center}
\includegraphics[scale=0.50]{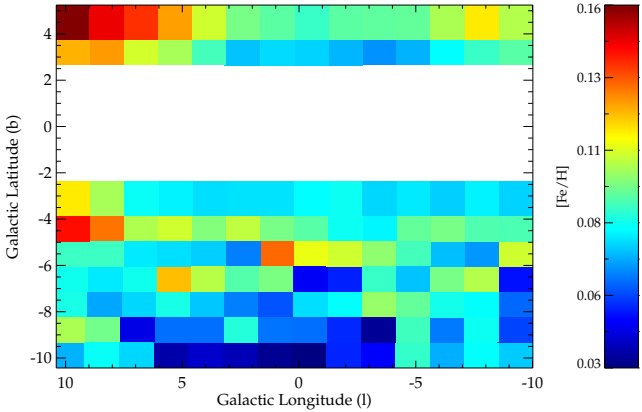}
\caption{The metallicity dispersion $\sigma_{dis}$ of each line of sight given by the individual VVV tiles, after varying the mean $K_s$ magnitude for 600 random values within 2$\sigma$ of the RC Gaussian fit.}
\label{mdf_dist}
\end{center}
\end{figure}

Bulge stars are often assumed to be well represented by a population at $\sim$8 kpc from the Sun \citep[e.g.][]{rattenbury07}. However, for a wide field study such as the one presented here, we must include two factors into our analysis: the bar inclination angle and the distance spread of stars.
As explained in Section~3, we have used the red clump method to derive the mean distance of the stars towards each line of sight, therefore the inclination angle of the bulge has been fully taken into consideration. However, by assuming that all stars are located at the same distance, corresponding to the distance modulus derived using the red clump magnitude, we are introducing an error which is directly related to the magnitude width of the red clump in the luminosity function. 
The width of the RC luminosity function, for a population located at the same distance, will be determined by the intrinsic width given by properties such as metallicity and age. However, the dominant factor in the case of the bulge RC magnitude distribution width will be the distance spread. We have therefore used a boothstrapping method in order to estimate the uncertainty arising from adopting a single mean magnitude and neglecting the width of the bulge RC. We re-calculated the mean metallicity after varying the mean magnitude of the red clump by a random value, within 2$\sigma$ of the mean RC $K_s$ magnitude. In each case, $\sigma$ is obtained from the gaussian fit to the luminosity function in Equation 1. We repeated this procedure 600 times for each VVV tile, since one tile corresponds to our resolution element used to derive the mean red clump magnitude. Figure~\ref{mdf_dist} shows the dispersion ($\sigma_{dis}$) in the mean metallicity obtained for each VVV tile after the iteration procedure.
 
\subsection{Residual differential extinction}

Another source of uncertainties in our derivation of the mean metallicity is due to a possible presence of residual differential extinction. Although a great improvement is provided by the high-resolution extinction map presented in \citet{gonzalez12}, the resolution of this map is limited by requiring a sufficient number of stars to fit the RC color distribution. The corresponding resolution of the map, under this limitation, was given by 2', 4' and 6' arcmin depending on latitude. Variations on scales smaller than these resolution elements are still possible, which then produce an increase in the width of the RC color and, therefore also in the RGB region that we use to derive the metallicities. As the reddening becomes larger towards the inner regions of the bulge, the residual reddening follows the same behavior. This creates an excess of redder stars in the CMD which would end up in a derivation of higher photometric metallicities for those stars and creating a metallicity gradient as the one we observe in our maps. Thus, we must ensure that the gradient we observe is not a product of differential reddening but a real feature.

We therefore evaluate the effect of residual differential extinction by studying the width of the $(J-K_s)_0$ color distribution of the RC, measured using a Gaussian fit to the RC color in each resolution element of the metallicity map. The dispersion, $\sigma_{(J-K_s)}$ of the Gaussian fit to RC color distribution will depend primarily on metallicity dispersion ($\sigma_{(J-K_s),Fe}$) as well as on the presence of differential reddening ($\sigma_{(J-K_s),red}$). We can thus parametrize $\sigma_{(J-K_s)}$ as $\sigma_{(J-K_s)}=\sqrt{\sigma_{(J-K_s),0}^2+\sigma_{(J-K_s),Fe}^2+\sigma_{(J-K_s),red}^2}$, where $\sigma_{(J-K_s),0}$ is the uniform intrisic width of the bulge RC population. In the context of this test we can then adopt the hypothesis of having neither a metallicity nor a metallicity dispersion gradient in the bulge, where both $\sigma_{(J-K_s),0}$ and $\sigma_{(J-K_s),Fe}$ can be assumed to be constant across the different regions and the single effect of residual differential extinction can be investigated. Then, under this assumption, the squared root difference between the $\sigma_{(J-K_s)}$ value in any region and that of the outermost region of our map, where the effects of extinction and certainly residual reddening are not present, would be a direct measurment of $\sigma_{(J-K_s),red}$. We calculated the $\sigma_{(J-K_s),red}$ values for each field using VVV field b217 as our reddening free control field.

\begin{figure}[htb]
\begin{center}
\includegraphics[scale=0.50]{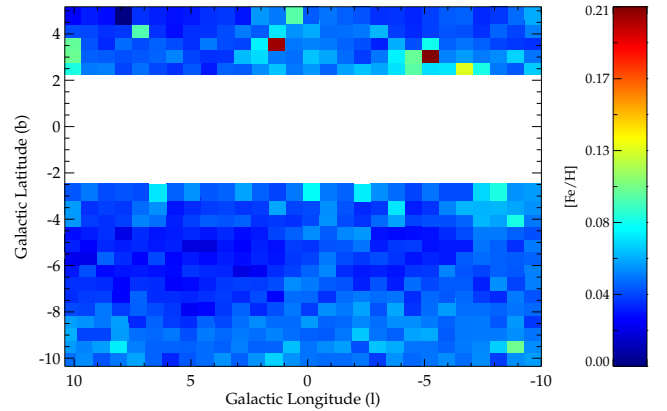}
\caption{Error in the mean metallicity derivation from residual differental reddening. Error values are given by the difference between the original [Fe/H] of the control field (b217) and the re-calculated metallicity after applying the $\sigma_{(J-K_s),red}$ corresponding to each field.}
\label{mdf_sigma}
\end{center}
\end{figure}

Finally, we applied a 1$\sigma_{(J-K_s),red}$ variation to the color of the RGB stars of our control field in the outer bulge and re-derived the metallicity distribution for the $\sigma_{(J-K_s),red}$ corresponding to each field. Figure~\ref{mdf_sigma} shows the map of $\sigma_{Fe,red}$ given as the difference between the original [Fe/H] of the control field and the new value, affected by each $\sigma_{(J-K_s),red}$. We observe that the residual differential reddening does not \textit{mask} the shape of the metallicity gradient and, most importantly, the error values in metallicity due to differential reddening are almost neglible for most regions (less than 0.05).

\subsection{Extinction law variations}
The reddening $E(J-K_s)$ from the map presented in Paper II can be converted to the corresponding extinctions $A_J$, $A_H$, and $A_{K_s}$ by adopting a corresponding extinction law. Most of the studies towards the Galactic bulge adopt standard extinction laws consistent with a $R_V\sim$3.1 such as \citet{cardelli89}. However, there are evidences for a non-standard extinction law, both in the optical and near-IR, towards the bulge. \citet{gould01} and \citet{udalski03} demonstrated that the optical reddening law toward the inner Galaxy is described by smaller total-to-selective ratios than the standard values measured for the local interstellar medium. \citet{udalski03} measured a value of $dA_I=dE(V-I)=1.1$ towards several bulge fields using OGLE photometry, much smaller than the $dA_I=dE(V-I)=1.4$ suggested by the standard interstellar extinction curve of $R_V$= 3:1 \citep{cardelli89, odonnell94}. Furthermore, not only was the optical reddening law toward the bulge found to be non-standard, it was also found to be
rapidly varying between sightlines \citep{nataf12}. The steeper extinction law toward the inner Galaxy has been subsequently confirmed with observations in the optical \citep{Revni10}, by analysis of RR Lyrae stars in OGLE-III \citep{pietru12}, and also in the near-IR \citep{nishi09, gosling09, schoedel10}.

\begin{figure}[htb]
\begin{center}
\includegraphics[scale=0.50]{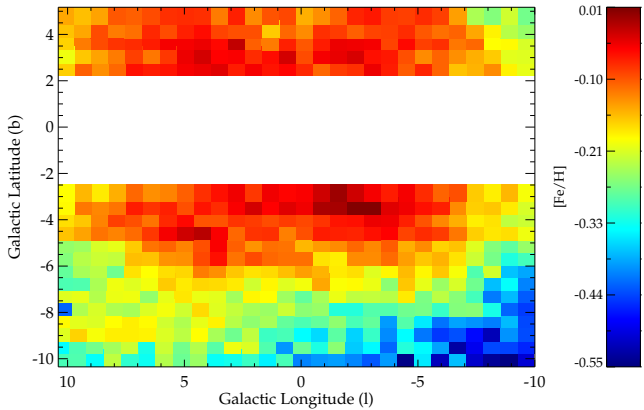}
\caption{Metallicity map for the Galactic bulge, as shown in Fig~\ref{mdf_full}, using the extinction law from Nishiyama et al. (2009)}
\label{mdf_full_nishi}
\end{center}
\end{figure}

\begin{figure}[htb]
\begin{center}
\includegraphics[scale=0.50]{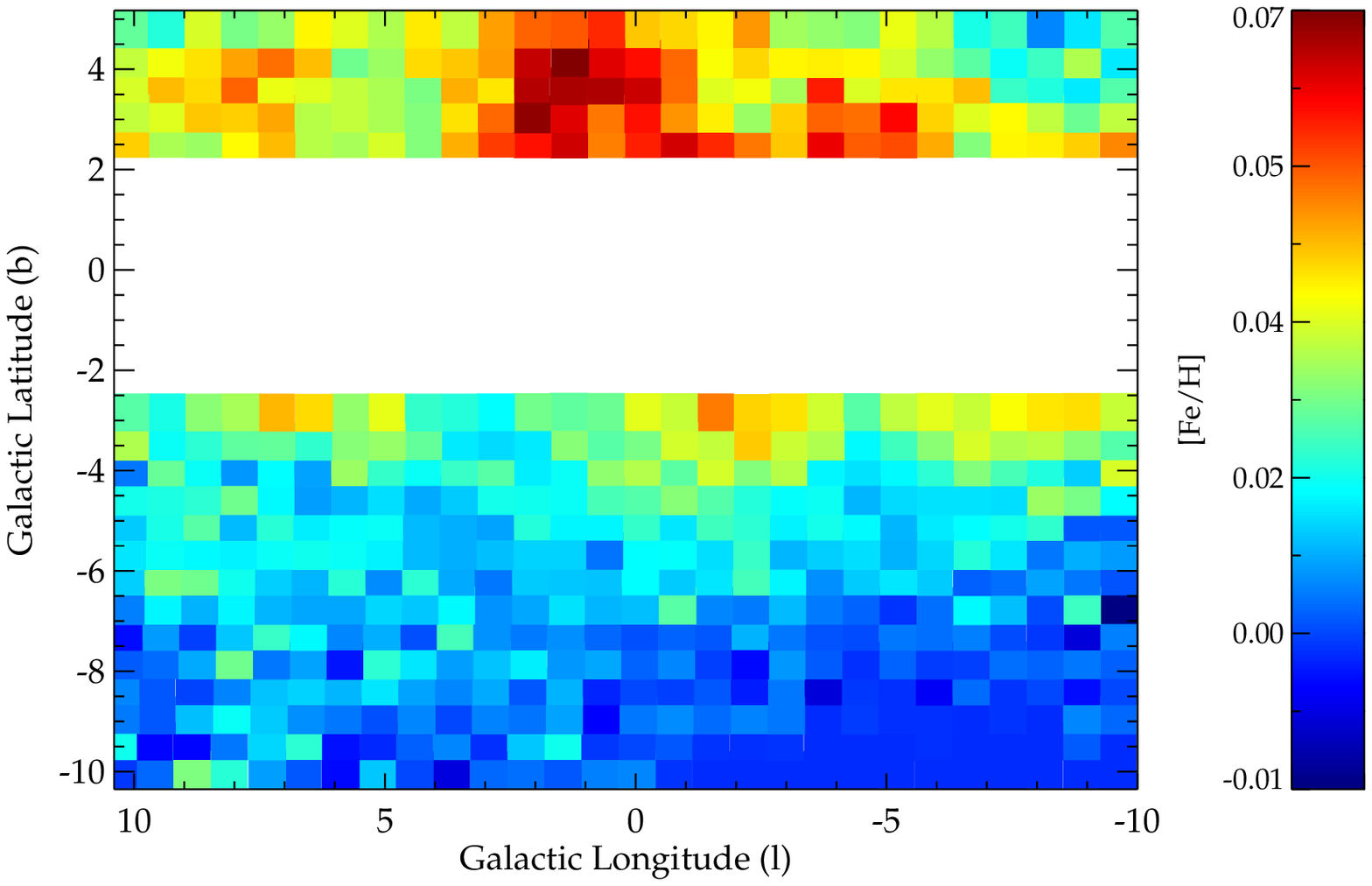}
\caption{Difference, $\rm [Fe/H]_{nishiyama}-[Fe/H]_{cardelli}$, between the metallicities derived using the extinction law of \citet{nishi09} and that of \citet{cardelli89} for the Galactic bulge.}
\label{mdf_law}
\end{center}
\end{figure}

An error in our metallicity calculations can be produced when adopting a given extinction law for the complete bulge region when such a law might no longer be valid in all regions. In particular, if a difference in the extinction law is present between the outer and inner bulge, a metallicity gradient could also be a consequence of this systematic error instead of a real feature. Therefore, we have derived our metallicity map using the standard law of \citet{cardelli89}, shown in Fig.~\ref{mdf_full}, as well as using that of \citet{nishi09} which is now shown in Fig.~\ref{mdf_full_nishi}. The difference between the metallicity obtained using both reddening laws allow us to evaluate its effect on the observed gradient. Figure~\ref{mdf_law} shows the difference between the metallicity map derived using \citet{cardelli89} and that of \citet{nishi09}. The maximum difference is observed to be of 0.10 dex in the mean metallicity and, as expected, it correlates with the amount of extinction. The variations in extinction law are therefore not enough to produce the observed gradient and it also does not follow the same spatial behavior of the metallicity gradient.

\begin{figure}[htb]
\begin{center}
\includegraphics[scale=0.60]{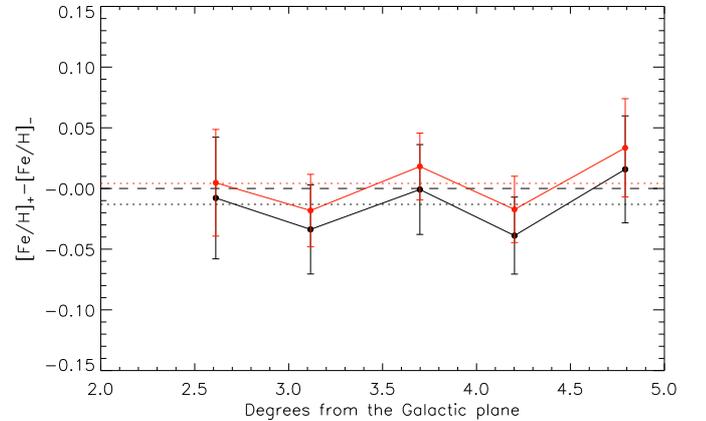}
\caption{Difference in the mean metallicity at different latitudes above ([Fe/H]$_{+}$) and below ([Fe/H]$_{-}$) the Galactic plane. The mean difference is shown as a doted line. Results obtained using the \citet{cardelli89} extinction law are shown in black and in red for those where the Nishiyama et al. (2009) law is adopted.}
\label{mdf_sym}
\end{center}
\end{figure}

Additionally, we can evaluate the possible effect of extinctions law variations, in our derived photometric metallicities. For this we must first assume that the metallicity distribution function is symmetric with respect to the Galactic plane. If this is the case, then the difference between metallicities measured at positive and negative latitudes should be zero. On the other hand, since the amount of reddening is not necessarily symmetric with respect to the plane, and it is in fact observed to be higher at positive latitudes than at negative ones \citep{marshall06, gonzalez12}, an asymmetric metallicity map would result if reddening and metallicity are not being properly disentagled because of, for example, the use of a non-suitable extinction law. Figure~\ref{mdf_sym} shows the mean difference found between metallicities above ($b>0$) and below ($b<0$) the Galactic plane as a function of distance from the plane. Error bars show the dispersion in this difference in metallicity for each latitude bin. 

We found a small systematic shift from zero ($\Delta$[Fe/H$]\sim-0.01$), in the mean metallicity difference, when using the \citet{cardelli89} extinction law, which is no longer present when the Nishiyama et al. (2009) law is adopted. This particular analysis would be independent of the adopted extinction law only if the amount of reddening was the same at both positive and negative latitudes. Therefore, we can interpret this result as the combined effect of the large amount of reddening at positive latitudes and the Cardelli et al. law being less adequate to correct for extinction effects in these inner Bulge regions. This effect is already included in our previously estimated errors for each resolution element of our metallicity map but we can now confirm the actual improvement of using the Nishiyama et al. (2009) extinction law in the inner Bulge regions ($b<5$).

\subsection{The metallicity error map}

After evaluating the individual sources of errors from the distance uncertainty, differential reddening, and the extinction law, we can obtain a map with the corresponding error for each of the subfields in our metallicity map, calculated as:

\begin{equation}
\sigma_{Fe}=\sqrt{\sigma_{Fe,dist}^2+\sigma_{Fe,red}^2+\sigma_{Fe,law}^2}
\end{equation}

\begin{figure}[htb]
\begin{center}
\includegraphics[scale=0.50]{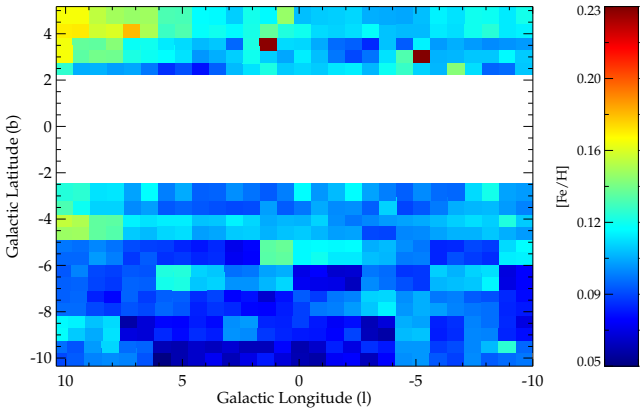}
\caption{Error map for the mean photometric metallicity for each subfield.}
\label{mdf_error}
\end{center}
\end{figure}

Figure~\ref{mdf_error} shows the error metallicity map for each subfield. This error does not include the \textit{method} error associated to obtain photometric metallicities. However, although the individual errors of photometric metallicities are expected to be larger than those from spectroscopy, we see differences lower than 0.05 dex between high resolution spectroscopic and photometric metallicities, when the mean value of the metallicity distribution is considered. We therefore consider this \textit{method} error in the mean to be neglictible compared to those previously addressed.

\section{The formation scenario of the bulge}

In the recent years, several authors have suggested that the shape of the metallicity distribution, and its variation across the bulge, is determined by the different contribution of two (or more) components. However, the origin of these components, as well as their main characteristics such as their age and metallicity, are not yet clear. Therefore, the complete view of the metallicity gradients accross the bulge is of high importance for constraining Galactic models. 



The presence of a radial metallicity gradient is commonly considered as a strong argument in favor of the classical bulge scenario. In our complete metallicity map we observe such gradient: a [Fe/H]$\sim$0.28 dex/kpc, value that is in remarkable agreement with the predicted gradient for bulge formed via dissipational collapse in the chemo-dynamical model discussed in \citet{grieco12}. 

However, structural and kinematical studies hint towards a different scenario. There is sufficient evidence for the Milky Way bulge structure to be consistent with a boxy-bulge, formed as a consequence of the vertical buckling of the bar. Several photometric studies have shown how the bar can be traced at latitudes well above the Galactic plane \citep{stanek97, lopez05, rattenbury07, dwek95, binney97}. In plane studies of the bar, in particular those based on near-IR star counts, have suggested the presence of a longer bar, in the plane of the Milky Way, which extends to larger longitudes and has a different orientation angle than the off-plane bar \citep{benjamin05, cabrera07, cabrera08, churchwell09}. \citet{martinez11} have recently presented a dynamical model, where a bar and boxy-bulge are formed through secular evolution, conciliating both long and thick bar components observed in the Milky Way into a single original structure.  Additionally, the observed cylindrical rotation of this structure led \citet{shen10} to suggest that no additional classical component is required to reproduce the Milky Way bulge kinematical properties. Furthermore, \citet{mcwilliam10} and \citet{nataf10} presented the observations of a double RC feature in the luminosity function of the bulge and interpreted it as two concentration of stars at different distances. \citet{mcwilliam10} suggested that this is the result of the bulge being X-shaped. This finding was later followed by a more complete mapping of the double RC in the work of \citet{saito11}, who provided a more general view of the X-shape morphology of the bulge. 

These conclusions regarding the morphology and kinematics of the Galactic bulge are consistent with results from simulations where a bar suffering from buckling instabilities results in the formation of a peanut or X-shaped structure \citep{athana05, debattista04, debattista06, martinez06} that puffs up from the plane of the galaxy. When these vertically thickened  bars are observed with an intermediate angle orientation, they would appear as boxy-bulges \citep{kuijken95,bureau_freeman99,patsis02,bureau06}. Additionally, we see from the metallicity map in Fig.~\ref{mdf_full} that stars at larger latitudes are more metal-rich at positive longitudes with respect to negative ones. Following the discussion in \citep{saito11}, at a given latitude, stars at negative longitudes are further away from us, because of the bar viewing angle, and therefore their true vertical distance from the plane will be larger than for stars at the near side of the bar. The observed horizontal pattern in the metallicity map of Fig.~2 would then be fully explained by the vertical metallicity gradient along the boxy-bulge. 

In order to reconcile these chemical and structural properties of the Galactic bulge, we first need to understand whether boxy-bulges alone can also show a metallicity gradient as the one observed in our map. \citet{bekki11} discussed this possibility, based on a Galactic model where a boxy-bulge is formed out of a two component disk. In their simulation, they are able to reproduce a vertical metallicity gradient when the bulge is secularly born from an original disk with its own vertical metallicity gradient, i.e. a metal-poor thick and metal-rich thin disk component, such as those observed in the Milky Way. The initial chemical properties of the stars in the disk would then produce the observed metallicity gradient, as the metal-poor stars that dominate the dinamically hotter thick disk would not be influenced by the mixing within the bar. 


The boxy-bulge model of \citet{martinez11}, which has shown to succesfully reproduce the stellar count profiles for both the inner and outer bulge, also shows a remarkable agreement with the complete metallicity gradient shown in this work (Martinez-Valpuesta et al. 2013, in preparation). We recall that the model does not include a classical component, hence demonstrating that the buckling process of a secularly evolved bar could be a sufficient process to explain the formation history of the bulge.

In support of the above scenario, there is the chemical similarity observed between bulge and disk stars. Metal-rich stars of the bulge are observed to be as alpha-poor as those of the thin disk, while metal-poor bulge stars show an alpha-element enhancement which is indistinguishable from that of the Galactic thick disk and dominate at higher scale heights \citep{melendez08, alves10, gonzalez11_alphas}. Furthermore, the mainly old stellar ages observed in the bulge \citep{zoccali03, clarkson11} place additional constrains to such formation scenario. The first original thick disk would have to be necessarily formed fast ($\sim$1 Gyr), most likely as the result of dissipative collapse or the merging of gas-rich subgalactic clumps, in order to be consistent with the early formation of the hotter, old and alpha-enhanced component. The inner stellar disk would then need to proceed with its star formation, relatively fast compared to the outer thin disk, to then become the inner parts of the boxy-bulge.   

Finally, we recall that recent studies have also suggested that the chemodynamical properties of the bulge are consistent with a change of the dominating population between the outer and inner bulge: a metal-poor, alpha-rich component with spheroid kinematics which is homogeneous along the minor axis, and a metal-rich, alpha-poor bar population dominating the inner bulge \citep{babusiaux10,hill11,gonzalez11_alphas}. Under this interpretation the vertical metallicity gradient will be naturally produced by the density distribution of both components at different lines of sight. Although this is also consistent with the general view seen in our complete metallicity map, the large individual metallicity errors in our photometric method in comparison to those from spectroscopy, do not allow to invenstigate further the bimodality in the metallicity distribution (see Fig.~1). Therefore, the presence of a bimodal metallicity distribution, and its correlation with kinematics, remains to be spectroscopically confirmed within a large spatial coverage.

\section{Conclusions}

We used the RC stars of the Galactic bulge as a standard candle to build a reddening map and determine distances. This allowed us to construct the color magnitude diagrams of bulge stars in the absolute plane ($M_{K_s}$,$(J-K_s)_0$) and then to obtain individual photometric metallicities for RGB stars in the bulge through interpolation on the RGB ridge lines of cluster sample from \citet[][]{valenti07}, taking into account the population corrections for the RC absolute magnitude as a function of metallicity, in an iterative procedure. The final metallicity distributions were compared to those obtained in the spectroscopic studies showing a remarkable agreement. The main result is the first high resolution metallicity map of the Milky Way bulge with a global coverage from $-10 < l < +10$ and $-10< b < +5$, excluding the central area $(|b|<2)$ where too high crowding and strong differential extinction compromise the measurements.

The global bulge metallicity map is consistent with a vertical metallicity gradient ($\sim0.04$ dex/deg), with metal-rich stars ([Fe/H]$\sim$0) dominating the inner bulge in regions closer to the galactic plane ($|b|<5$). At larger scale heights, the mean metallicity of the bulge population becomes more metal-poor. This fits in the scenario of older, metal-poor stars dominating the bulge at higher scale heights above the plane as a consequence of original gradients present in the Galactic disk. Therefore our metallicity map supports the formation of the boxy bulge from the buckling instabillity of the Galactic bar \citep[][]{bekki11, martinez11}. This scenario is also supported by the chemical similarities observed between the metal-poor bulge and the thick disk stars.

\begin{acknowledgements}
We acknowgledge the referee, David Nataf, for a carefull review and valuable comments which contributed to the improvement of the article. We warmly thank Alvio Renzini and Vanessa Hill for providing useful comments and suggestions to this work. We are grateful to Inma Martinez-Valpuesta and Ortwin Gerhard for sharing their model results before publication.  We gratefully acknowledge use of data from the ESO Public Survey program ID 179.B-2002 taken with the VISTA telescope, data products from the Cambridge Astronomical Survey Unit. We acknowledge funding from the FONDAP Center for Astrophysics 15010003, the BASAL CATA Center for Astrophysics and Associated Technologies PFB-06, the MILENIO Milky Way Millennium Nucleus from the Ministry of Economycs ICM grant P07-021-F and Proyectos FONDECYT Regular 1110393 and 1090213. MZ is also partially supported by Proyecto Anillo ACT-86. This work was co-funded under
the Marie Curie Actions of the European Commission (FP7-COFUND). This publication uses data products from the Two Micron All Sky Survey, which is a joint project of the University of Massachusetts and Infrared Processing and Analysis Center/California Institute of Technology, funded by the National Aeronautics and Space Administration and the National Science Foundation. We warmly thank the ESO Paranal Observatory staff for performing the observations. 
\end{acknowledgements}

\renewcommand*{\bibfont}{\tiny}
\bibliographystyle{aa}

\bibliography{mybiblio}

\begin{thebibliography}{90}
\expandafter\ifx\csname natexlab\endcsname\relax\def\natexlab#1{#1}\fi

\bibitem[{{Alves-Brito} {et~al.}(2010){Alves-Brito}, {Mel{\'e}ndez}, {Asplund},
  {Ram{\'{\i}}rez}, \& {Yong}}]{alves10}
{Alves-Brito}, A., {Mel{\'e}ndez}, J., {Asplund}, M., {Ram{\'{\i}}rez}, I., \&
  {Yong}, D. 2010, \aap, 513, A35+

\bibitem[{{Athanassoula}(2005)}]{athana05}
{Athanassoula}, E. 2005, \mnras, 358, 1477

\bibitem[{{Athanassoula} \& {Misiriotis}(2002)}]{athana02}
{Athanassoula}, E. \& {Misiriotis}, A. 2002, \mnras, 330, 35

\bibitem[{{Babusiaux} {et~al.}(2010){Babusiaux}, {G{\'o}mez}, {Hill}, {Royer},
  {Zoccali}, {Arenou}, {Fux}, {Lecureur}, {Schultheis}, {Barbuy}, {Minniti}, \&
  {Ortolani}}]{babusiaux10}
{Babusiaux}, C., {G{\'o}mez}, A., {Hill}, V., {et~al.} 2010, \aap, 519, A77+

\bibitem[{{Bekki} \& {Tsujimoto}(2011)}]{bekki11}
{Bekki}, K. \& {Tsujimoto}, T. 2011, \mnras, 416, L60

\bibitem[{{Benjamin} {et~al.}(2005){Benjamin}, {Churchwell}, {Babler},
  {Indebetouw}, {Meade}, {Whitney}, {Watson}, {Wolfire}, {Wolff}, {Ignace},
  {Bania}, {Bracker}, {Clemens}, {Chomiuk}, {Cohen}, {Dickey}, {Jackson},
  {Kobulnicky}, {Mercer}, {Mathis}, {Stolovy}, \& {Uzpen}}]{benjamin05}
{Benjamin}, R.~A., {Churchwell}, E., {Babler}, B.~L., {et~al.} 2005, \apjl,
  630, L149

\bibitem[{{Bensby} {et~al.}(2012){Bensby}, {Feltzing}, {Gould}, {Johnson},
  {Asplund}, {Ad{\'e}n}, {Mel{\'e}ndez}, {Cohen}, {Thompson}, {Lucatello}, \&
  {Gal-Yam}}]{bensby12}
{Bensby}, T., {Feltzing}, S., {Gould}, A., {et~al.} 2012, in Astronomical
  Society of the Pacific Conference Series, Vol. 458, Galactic Archaeology:
  Near-Field Cosmology and the Formation of the Milky Way, ed. W.~{Aoki},
  M.~{Ishigaki}, T.~{Suda}, T.~{Tsujimoto}, \& N.~{Arimoto}, 203

\bibitem[{{Binney} {et~al.}(1997){Binney}, {Gerhard}, \& {Spergel}}]{binney97}
{Binney}, J., {Gerhard}, O., \& {Spergel}, D. 1997, \mnras, 288, 365

\bibitem[{{Bureau} {et~al.}(2006){Bureau}, {Aronica}, {Athanassoula},
  {Dettmar}, {Bosma}, \& {Freeman}}]{bureau06}
{Bureau}, M., {Aronica}, G., {Athanassoula}, E., {et~al.} 2006, \mnras, 370,
  753

\bibitem[{{Bureau} \& {Freeman}(1999)}]{bureau_freeman99}
{Bureau}, M. \& {Freeman}, K.~C. 1999, \aj, 118, 126

\bibitem[{{Cabrera-Lavers} {et~al.}(2008){Cabrera-Lavers},
  {Gonz{\'a}lez-Fern{\'a}ndez}, {Garz{\'o}n}, {Hammersley}, \&
  {L{\'o}pez-Corredoira}}]{cabrera08}
{Cabrera-Lavers}, A., {Gonz{\'a}lez-Fern{\'a}ndez}, C., {Garz{\'o}n}, F.,
  {Hammersley}, P.~L., \& {L{\'o}pez-Corredoira}, M. 2008, \aap, 491, 781

\bibitem[{{Cabrera-Lavers} {et~al.}(2007){Cabrera-Lavers}, {Hammersley},
  {Gonz{\'a}lez-Fern{\'a}ndez}, {L{\'o}pez-Corredoira}, {Garz{\'o}n}, \&
  {Mahoney}}]{cabrera07}
{Cabrera-Lavers}, A., {Hammersley}, P.~L., {Gonz{\'a}lez-Fern{\'a}ndez}, C.,
  {et~al.} 2007, \aap, 465, 825

\bibitem[{{Cardelli} {et~al.}(1989){Cardelli}, {Clayton}, \&
  {Mathis}}]{cardelli89}
{Cardelli}, J.~A., {Clayton}, G.~C., \& {Mathis}, J.~S. 1989, \apj, 345, 245

\bibitem[{{Carollo} {et~al.}(2007){Carollo}, {Scarlata}, {Stiavelli}, {Wyse},
  \& {Mayer}}]{carollo07}
{Carollo}, C.~M., {Scarlata}, C., {Stiavelli}, M., {Wyse}, R.~F.~G., \&
  {Mayer}, L. 2007, \apj, 658, 960

\bibitem[{{Churchwell} {et~al.}(2009){Churchwell}, {Babler}, {Meade},
  {Whitney}, {Benjamin}, {Indebetouw}, {Cyganowski}, {Robitaille}, {Povich},
  {Watson}, \& {Bracker}}]{churchwell09}
{Churchwell}, E., {Babler}, B.~L., {Meade}, M.~R., {et~al.} 2009, \pasp, 121,
  213

\bibitem[{{Clarkson} {et~al.}(2011){Clarkson}, {Sahu}, {Anderson}, {Rich},
  {Smith}, {Brown}, {Bond}, {Livio}, {Minniti}, {Renzini}, \&
  {Zoccali}}]{clarkson11}
{Clarkson}, W.~I., {Sahu}, K.~C., {Anderson}, J., {et~al.} 2011, \apj, 735, 37

\bibitem[{{Combes} {et~al.}(1990){Combes}, {Debbasch}, {Friedli}, \&
  {Pfenniger}}]{combes90}
{Combes}, F., {Debbasch}, F., {Friedli}, D., \& {Pfenniger}, D. 1990, \aap,
  233, 82

\bibitem[{{Combes} \& {Sanders}(1981)}]{combes81}
{Combes}, F. \& {Sanders}, R.~H. 1981, \aap, 96, 164

\bibitem[{{Daddi} {et~al.}(2010){Daddi}, {Bournaud}, {Walter}, {Dannerbauer},
  {Carilli}, {Dickinson}, {Elbaz}, {Morrison}, {Riechers}, {Onodera}, {Salmi},
  {Krips}, \& {Stern}}]{daddi10}
{Daddi}, E., {Bournaud}, F., {Walter}, F., {et~al.} 2010, \apj, 713, 686

\bibitem[{{Debattista} {et~al.}(2006){Debattista}, {Mayer}, {Carollo}, {Moore},
  {Wadsley}, \& {Quinn}}]{debattista06}
{Debattista}, V.~P., {Mayer}, L., {Carollo}, C.~M., {et~al.} 2006, \apj, 645,
  209

\bibitem[{{Debattista} \& {Sellwood}(1998)}]{debattista98}
{Debattista}, V.~P. \& {Sellwood}, J.~A. 1998, \apjl, 493, L5

\bibitem[{{Debattista} \& {Williams}(2004)}]{debattista04}
{Debattista}, V.~P. \& {Williams}, T.~B. 2004, \apj, 605, 714

\bibitem[{{Dwek} {et~al.}(1995){Dwek}, {Arendt}, {Hauser}, {Kelsall}, {Lisse},
  {Moseley}, {Silverberg}, {Sodroski}, \& {Weiland}}]{dwek95}
{Dwek}, E., {Arendt}, R.~G., {Hauser}, M.~G., {et~al.} 1995, \apj, 445, 716

\bibitem[{{Elmegreen} {et~al.}(2008){Elmegreen}, {Bournaud}, \&
  {Elmegreen}}]{elmegreen08}
{Elmegreen}, B.~G., {Bournaud}, F., \& {Elmegreen}, D.~M. 2008, \apj, 688, 67

\bibitem[{{Emsellem} {et~al.}(2004){Emsellem}, {Cappellari}, {Peletier},
  {McDermid}, {Bacon}, {Bureau}, {Copin}, {Davies}, {Krajnovi{\'c}},
  {Kuntschner}, {Miller}, \& {de Zeeuw}}]{emsellem04}
{Emsellem}, E., {Cappellari}, M., {Peletier}, R.~F., {et~al.} 2004, \mnras,
  352, 721

\bibitem[{{Eskridge} {et~al.}(2000){Eskridge}, {Frogel}, {Pogge}, {Quillen},
  {Davies}, {DePoy}, {Houdashelt}, {Kuchinski}, {Ram{\'{\i}}rez}, {Sellgren},
  {Terndrup}, \& {Tiede}}]{eskridge00}
{Eskridge}, P.~B., {Frogel}, J.~A., {Pogge}, R.~W., {et~al.} 2000, \aj, 119,
  536

\bibitem[{{F{\"o}rster Schreiber} {et~al.}(2009){F{\"o}rster Schreiber},
  {Genzel}, {Bouch{\'e}}, {Cresci}, {Davies}, {Buschkamp}, {Shapiro},
  {Tacconi}, {Hicks}, {Genel}, {Shapley}, {Erb}, {Steidel}, {Lutz},
  {Eisenhauer}, {Gillessen}, {Sternberg}, {Renzini}, {Cimatti}, {Daddi},
  {Kurk}, {Lilly}, {Kong}, {Lehnert}, {Nesvadba}, {Verma}, {McCracken},
  {Arimoto}, {Mignoli}, \& {Onodera}}]{fs09}
{F{\"o}rster Schreiber}, N.~M., {Genzel}, R., {Bouch{\'e}}, N., {et~al.} 2009,
  \apj, 706, 1364

\bibitem[{{Freeman}(2008)}]{freeman08}
{Freeman}, K.~C. 2008, in IAU Symposium, Vol. 245, IAU Symposium, ed.
  {M.~Bureau, E.~Athanassoula, \& B.~Barbuy}, 3--10

\bibitem[{{Gadotti}(2009)}]{gadotti09}
{Gadotti}, D.~A. 2009, \mnras, 393, 1531

\bibitem[{{Gadotti}(2011)}]{gadotti11}
{Gadotti}, D.~A. 2011, ArXiv:1101.2714

\bibitem[{{Gadotti} \& {de Souza}(2006)}]{gadotti06}
{Gadotti}, D.~A. \& {de Souza}, R.~E. 2006, \apjs, 163, 270

\bibitem[{{Genzel} {et~al.}(2008){Genzel}, {Burkert}, {Bouch{\'e}}, {Cresci},
  {F{\"o}rster Schreiber}, {Shapley}, {Shapiro}, {Tacconi}, {Buschkamp},
  {Cimatti}, {Daddi}, {Davies}, {Eisenhauer}, {Erb}, {Genel}, {Gerhard},
  {Hicks}, {Lutz}, {Naab}, {Ott}, {Rabien}, {Renzini}, {Steidel}, {Sternberg},
  \& {Lilly}}]{genzel08}
{Genzel}, R., {Burkert}, A., {Bouch{\'e}}, N., {et~al.} 2008, \apj, 687, 59

\bibitem[{{Genzel} {et~al.}(2006){Genzel}, {Tacconi}, {Eisenhauer},
  {F{\"o}rster Schreiber}, {Cimatti}, {Daddi}, {Bouch{\'e}}, {Davies},
  {Lehnert}, {Lutz}, {Nesvadba}, {Verma}, {Abuter}, {Shapiro}, {Sternberg},
  {Renzini}, {Kong}, {Arimoto}, \& {Mignoli}}]{genzel06}
{Genzel}, R., {Tacconi}, L.~J., {Eisenhauer}, F., {et~al.} 2006, 442, 786

\bibitem[{{Gonzalez} {et~al.}(2011{\natexlab{a}}){Gonzalez}, {Rejkuba},
  {Zoccali}, {Hill}, {Battaglia}, {Babusiaux}, {Minniti}, {Barbuy},
  {Alves-Brito}, {Renzini}, {Gomez}, \& {Ortolani}}]{gonzalez11_alphas}
{Gonzalez}, O.~A., {Rejkuba}, M., {Zoccali}, M., {et~al.} 2011{\natexlab{a}},
  \aap, 530, A54+

\bibitem[{{Gonzalez} {et~al.}(2011{\natexlab{b}}){Gonzalez}, {Rejkuba},
  {Zoccali}, {Valenti}, \& {Minniti}}]{gonzalez11_I}
{Gonzalez}, O.~A., {Rejkuba}, M., {Zoccali}, M., {Valenti}, E., \& {Minniti},
  D. 2011{\natexlab{b}}, \aap, 534, A3

\bibitem[{{Gonzalez} {et~al.}(2012){Gonzalez}, {Rejkuba}, {Zoccali}, {Valenti},
  {Minniti}, {Schultheis}, {Tobar}, \& {Chen}}]{gonzalez12}
{Gonzalez}, O.~A., {Rejkuba}, M., {Zoccali}, M., {et~al.} 2012, \aap, 543, A13

\bibitem[{{Gosling} {et~al.}(2009){Gosling}, {Bandyopadhyay}, \&
  {Blundell}}]{gosling09}
{Gosling}, A.~J., {Bandyopadhyay}, R.~M., \& {Blundell}, K.~M. 2009, \mnras,
  394, 2247

\bibitem[{{Gould} {et~al.}(2001){Gould}, {Stutz}, \& {Frogel}}]{gould01}
{Gould}, A., {Stutz}, A., \& {Frogel}, J.~A. 2001, \apj, 547, 590

\bibitem[{{Grieco} {et~al.}(2012){Grieco}, {Matteucci}, {Pipino}, \&
  {Cescutti}}]{grieco12}
{Grieco}, V., {Matteucci}, F., {Pipino}, A., \& {Cescutti}, G. 2012, \aap, 548,
  A60

\bibitem[{{Hill} {et~al.}(2011){Hill}, {Lecureur}, {Gomez}, {Zoccali},
  {Schultheis}, {Babusiaux}, {Royer}, {Barbuy}, {Arenou}, {Minniti}, \&
  {Ortolani}}]{hill11}
{Hill}, V., {Lecureur}, A., {Gomez}, A., {et~al.} 2011, \aap

\bibitem[{{Immeli} {et~al.}(2004){Immeli}, {Samland}, {Gerhard}, \&
  {Westera}}]{immeli04}
{Immeli}, A., {Samland}, M., {Gerhard}, O., \& {Westera}, P. 2004, \aap, 413,
  547

\bibitem[{{Jablonka} {et~al.}(2007){Jablonka}, {Gorgas}, \&
  {Goudfrooij}}]{jablonka07}
{Jablonka}, P., {Gorgas}, J., \& {Goudfrooij}, P. 2007, \aap, 474, 763

\bibitem[{{Jablonka} \& {Sarajedini}(2005)}]{jablonka05}
{Jablonka}, P. \& {Sarajedini}, A. 2005, in IAU Symposium, Vol. 228, From
  Lithium to Uranium: Elemental Tracers of Early Cosmic Evolution, ed.
  V.~{Hill}, P.~{Fran{\c c}ois}, \& F.~{Primas}, 525--530

\bibitem[{{Johnson} {et~al.}(2011){Johnson}, {Rich}, {Fulbright}, {Valenti}, \&
  {McWilliam}}]{johnson11}
{Johnson}, C.~I., {Rich}, R.~M., {Fulbright}, J.~P., {Valenti}, E., \&
  {McWilliam}, A. 2011, \apj, 732, 108

\bibitem[{{Kauffmann} {et~al.}(1999){Kauffmann}, {Colberg}, {Diaferio}, \&
  {White}}]{kauffmann99}
{Kauffmann}, G., {Colberg}, J.~M., {Diaferio}, A., \& {White}, S.~D.~M. 1999,
  \mnras, 303, 188

\bibitem[{{Kormendy} \& {Bruzual A.}(1978)}]{kormendy78}
{Kormendy}, J. \& {Bruzual A.}, G. 1978, \apjl, 223, L63

\bibitem[{{Kormendy} \& {Illingworth}(1982)}]{kormendy82}
{Kormendy}, J. \& {Illingworth}, G. 1982, \apj, 256, 460

\bibitem[{{Kormendy} \& {Kennicutt}(2004)}]{kormendy04}
{Kormendy}, J. \& {Kennicutt}, Jr., R.~C. 2004, 42, 603

\bibitem[{{Kuijken} \& {Merrifield}(1995)}]{kuijken95}
{Kuijken}, K. \& {Merrifield}, M.~R. 1995, \apjl, 443, L13

\bibitem[{{L{\'o}pez-Corredoira} {et~al.}(2005){L{\'o}pez-Corredoira},
  {Cabrera-Lavers}, \& {Gerhard}}]{lopez05}
{L{\'o}pez-Corredoira}, M., {Cabrera-Lavers}, A., \& {Gerhard}, O.~E. 2005,
  \aap, 439, 107

\bibitem[{{MacArthur} {et~al.}(2008){MacArthur}, {Ellis}, {Treu}, {U}, {Bundy},
  \& {Moran}}]{macarthur08}
{MacArthur}, L.~A., {Ellis}, R.~S., {Treu}, T., {et~al.} 2008, \apj, 680, 70

\bibitem[{{Marshall} {et~al.}(2006){Marshall}, {Robin}, {Reyl{\'e}},
  {Schultheis}, \& {Picaud}}]{marshall06}
{Marshall}, D.~J., {Robin}, A.~C., {Reyl{\'e}}, C., {Schultheis}, M., \&
  {Picaud}, S. 2006, \aap, 453, 635

\bibitem[{{Martinez-Valpuesta} \& {Gerhard}(2011)}]{martinez11}
{Martinez-Valpuesta}, I. \& {Gerhard}, O. 2011, \apjl, 734, L20+

\bibitem[{{Martinez-Valpuesta} {et~al.}(2006){Martinez-Valpuesta}, {Shlosman},
  \& {Heller}}]{martinez06}
{Martinez-Valpuesta}, I., {Shlosman}, I., \& {Heller}, C. 2006, \apj, 637, 214

\bibitem[{{McWilliam} \& {Zoccali}(2010)}]{mcwilliam10}
{McWilliam}, A. \& {Zoccali}, M. 2010, \apj, 724, 1491

\bibitem[{{Mel{\'e}ndez} {et~al.}(2008){Mel{\'e}ndez}, {Asplund},
  {Alves-Brito}, {Cunha}, {Barbuy}, {Bessell}, {Chiappini}, {Freeman},
  {Ram{\'{\i}}rez}, {Smith}, \& {Yong}}]{melendez08}
{Mel{\'e}ndez}, J., {Asplund}, M., {Alves-Brito}, A., {et~al.} 2008, \aap, 484,
  L21

\bibitem[{{Minniti} {et~al.}(2010){Minniti}, {Lucas}, {Emerson}, {Saito},
  {Hempel}, {Pietrukowicz}, {Ahumada}, {Alonso}, {Alonso-Garcia}, {Arias},
  {Bandyopadhyay}, {Barb{\'a}}, {Barbuy}, {Bedin}, {Bica}, {Borissova},
  {Bronfman}, {Carraro}, {Catelan}, {Clari{\'a}}, {Cross}, {de Grijs},
  {D{\'e}k{\'a}ny}, {Drew}, {Fari{\~n}a}, {Feinstein}, {Fern{\'a}ndez
  Laj{\'u}s}, {Gamen}, {Geisler}, {Gieren}, {Goldman}, {Gonzalez}, {Gunthardt},
  {Gurovich}, {Hambly}, {Irwin}, {Ivanov}, {Jord{\'a}n}, {Kerins}, {Kinemuchi},
  {Kurtev}, {L{\'o}pez-Corredoira}, {Maccarone}, {Masetti}, {Merlo},
  {Messineo}, {Mirabel}, {Monaco}, {Morelli}, {Padilla}, {Palma}, {Parisi},
  {Pignata}, {Rejkuba}, {Roman-Lopes}, {Sale}, {Schreiber}, {Schr{\"o}der},
  {Smith}, {Sodr{\'e}}, {Soto}, {Tamura}, {Tappert}, {Thompson}, {Toledo},
  {Zoccali}, \& {Pietrzynski}}]{vvv10}
{Minniti}, D., {Lucas}, P.~W., {Emerson}, J.~P., {et~al.} 2010, 15, 433

\bibitem[{{Moorthy} \& {Holtzman}(2006)}]{moorthy06}
{Moorthy}, B.~K. \& {Holtzman}, J.~A. 2006, \mnras, 371, 583

\bibitem[{{Nataf} {et~al.}(2012){Nataf}, {Gould}, {Fouqu{\'e}}, {Gonzalez},
  {Johnson}, {Skowron}, {Udalski}, {Szyma{\'n}ski}, {Kubiak},
  {Pietrzy{\'n}ski}, {Soszy{\'n}ski}, {Ulaczyk}, {Wyrzykowski}, \&
  {Poleski}}]{nataf12}
{Nataf}, D.~M., {Gould}, A., {Fouqu{\'e}}, P., {et~al.} 2012, arXiv:1208.1263

\bibitem[{{Nataf} {et~al.}(2010){Nataf}, {Udalski}, {Gould}, {Fouqu{\'e}}, \&
  {Stanek}}]{nataf10}
{Nataf}, D.~M., {Udalski}, A., {Gould}, A., {Fouqu{\'e}}, P., \& {Stanek},
  K.~Z. 2010, \apjl, 721, L28

\bibitem[{{Ness} {et~al.}(2012){Ness}, {Freeman}, {Athanassoula},
  {Wylie-De-Boer}, {Bland-Hawthorn}, {Lewis}, {Yong}, {Asplund}, {Lane},
  {Kiss}, \& {Ibata}}]{ness12}
{Ness}, M., {Freeman}, K., {Athanassoula}, E., {et~al.} 2012, \apj, 756, 22

\bibitem[{{Nishiyama} {et~al.}(2009){Nishiyama}, {Tamura}, {Hatano}, {Kato},
  {Tanab{\'e}}, {Sugitani}, \& {Nagata}}]{nishi09}
{Nishiyama}, S., {Tamura}, M., {Hatano}, H., {et~al.} 2009, \apj, 696, 1407

\bibitem[{{O'Donnell}(1994)}]{odonnell94}
{O'Donnell}, J.~E. 1994, \apj, 422, 158

\bibitem[{{Patsis} {et~al.}(2002){Patsis}, {Skokos}, \&
  {Athanassoula}}]{patsis02}
{Patsis}, P.~A., {Skokos}, C., \& {Athanassoula}, E. 2002, \mnras, 337, 578

\bibitem[{{Peletier} {et~al.}(1999){Peletier}, {Balcells}, {Davies},
  {Andredakis}, {Vazdekis}, {Burkert}, \& {Prada}}]{Peletier99}
{Peletier}, R.~F., {Balcells}, M., {Davies}, R.~L., {et~al.} 1999, \mnras, 310,
  703

\bibitem[{{Peng} {et~al.}(2010){Peng}, {Lilly}, {Kova{\v c}}, {Bolzonella},
  {Pozzetti}, {Renzini}, {Zamorani}, {Ilbert}, {Knobel}, {Iovino}, {Maier},
  {Cucciati}, {Tasca}, {Carollo}, {Silverman}, {Kampczyk}, {de Ravel},
  {Sanders}, {Scoville}, {Contini}, {Mainieri}, {Scodeggio}, {Kneib}, {Le
  F{\`e}vre}, {Bardelli}, {Bongiorno}, {Caputi}, {Coppa}, {de la Torre},
  {Franzetti}, {Garilli}, {Lamareille}, {Le Borgne}, {Le Brun}, {Mignoli},
  {Perez Montero}, {Pello}, {Ricciardelli}, {Tanaka}, {Tresse}, {Vergani},
  {Welikala}, {Zucca}, {Oesch}, {Abbas}, {Barnes}, {Bordoloi}, {Bottini},
  {Cappi}, {Cassata}, {Cimatti}, {Fumana}, {Hasinger}, {Koekemoer},
  {Leauthaud}, {Maccagni}, {Marinoni}, {McCracken}, {Memeo}, {Meneux}, {Nair},
  {Porciani}, {Presotto}, \& {Scaramella}}]{peng10}
{Peng}, Y., {Lilly}, S.~J., {Kova{\v c}}, K., {et~al.} 2010, \apj, 721, 193

\bibitem[{{P{\'e}rez} {et~al.}(2009){P{\'e}rez}, {S{\'a}nchez-Bl{\'a}zquez}, \&
  {Zurita}}]{perez09}
{P{\'e}rez}, I., {S{\'a}nchez-Bl{\'a}zquez}, P., \& {Zurita}, A. 2009, \aap,
  495, 775

\bibitem[{{Pietrinferni} {et~al.}(2004){Pietrinferni}, {Cassisi}, {Salaris}, \&
  {Castelli}}]{pietri04}
{Pietrinferni}, A., {Cassisi}, S., {Salaris}, M., \& {Castelli}, F. 2004, \apj,
  612, 168

\bibitem[{{Pietrukowicz} {et~al.}(2012){Pietrukowicz}, {Udalski},
  {Soszy{\'n}ski}, {Nataf}, {Wyrzykowski}, {Poleski}, {Koz{\l}owski},
  {Szyma{\'n}ski}, {Kubiak}, {Pietrzy{\'n}ski}, \& {Ulaczyk}}]{pietru12}
{Pietrukowicz}, P., {Udalski}, A., {Soszy{\'n}ski}, I., {et~al.} 2012, \apj,
  750, 169

\bibitem[{{Rattenbury} {et~al.}(2007){Rattenbury}, {Mao}, {Sumi}, \&
  {Smith}}]{rattenbury07}
{Rattenbury}, N.~J., {Mao}, S., {Sumi}, T., \& {Smith}, M.~C. 2007, \mnras,
  378, 1064

\bibitem[{{Renzini}(1998)}]{renzini98}
{Renzini}, A. 1998, \aj, 115, 2459

\bibitem[{{Renzini}(2009)}]{renzini09}
{Renzini}, A. 2009, \mnras, 398, L58

\bibitem[{{Revnivtsev} {et~al.}(2010){Revnivtsev}, {van den Berg}, {Burenin},
  {Grindlay}, {Karasev}, \& {Forman}}]{Revni10}
{Revnivtsev}, M., {van den Berg}, M., {Burenin}, R., {et~al.} 2010, \aap, 515,
  A49

\bibitem[{{Rich} {et~al.}(2007){Rich}, {Origlia}, \&
  {Valenti}}]{rich_origlia07}
{Rich}, R.~M., {Origlia}, L., \& {Valenti}, E. 2007, \apjl, 665, L119

\bibitem[{{Saito} {et~al.}(2012){Saito}, {Hempel}, {Minniti}, {Lucas},
  {Rejkuba}, {Toledo}, {Gonzalez}, {Alonso-Garc{\'{\i}}a}, {Irwin},
  {Gonzalez-Solares}, {Hodgkin}, {Lewis}, {Cross}, {Ivanov}, {Kerins},
  {Emerson}, {Soto}, {Am{\^o}res}, {Gurovich}, {D{\'e}k{\'a}ny}, {Angeloni},
  {Beamin}, {Catelan}, {Padilla}, {Zoccali}, {Pietrukowicz}, {Moni Bidin},
  {Mauro}, {Geisler}, {Folkes}, {Sale}, {Borissova}, {Kurtev}, {Ahumada},
  {Alonso}, {Adamson}, {Arias}, {Bandyopadhyay}, {Barb{\'a}}, {Barbuy},
  {Baume}, {Bedin}, {Bellini}, {Benjamin}, {Bica}, {Bonatto}, {Bronfman},
  {Carraro}, {Chen{\`e}}, {Clari{\'a}}, {Clarke}, {Contreras}, {Corvill{\'o}n},
  {de Grijs}, {Dias}, {Drew}, {Fari{\~n}a}, {Feinstein},
  {Fern{\'a}ndez-Laj{\'u}s}, {Gamen}, {Gieren}, {Goldman},
  {Gonz{\'a}lez-Fern{\'a}ndez}, {Grand}, {Gunthardt}, {Hambly}, {Hanson},
  {He{\l}miniak}, {Hoare}, {Huckvale}, {Jord{\'a}n}, {Kinemuchi}, {Longmore},
  {L{\'o}pez-Corredoira}, {Maccarone}, {Majaess}, {Mart{\'{\i}}n}, {Masetti},
  {Mennickent}, {Mirabel}, {Monaco}, {Morelli}, {Motta}, {Palma}, {Parisi},
  {Parker}, {Pe{\~n}aloza}, {Pietrzy{\'n}ski}, {Pignata}, {Popescu}, {Read},
  {Rojas}, {Roman-Lopes}, {Ruiz}, {Saviane}, {Schreiber}, {Schr{\"o}der},
  {Sharma}, {Smith}, {Sodr{\'e}}, {Stead}, {Stephens}, {Tamura}, {Tappert},
  {Thompson}, {Valenti}, {Vanzi}, {Walton}, {Weidmann}, \&
  {Zijlstra}}]{saito12}
{Saito}, R.~K., {Hempel}, M., {Minniti}, D., {et~al.} 2012, \aap, 537, A107

\bibitem[{{Saito} {et~al.}(2011){Saito}, {Zoccali}, {McWilliam}, {Minniti},
  {Gonzalez}, \& {Hill}}]{saito11}
{Saito}, R.~K., {Zoccali}, M., {McWilliam}, A., {et~al.} 2011, \aj, 142, 76

\bibitem[{{Salaris} \& {Girardi}(2002)}]{salaris+girardi02}
{Salaris}, M. \& {Girardi}, L. 2002, \mnras, 337, 332

\bibitem[{{Sch{\"o}del} {et~al.}(2010){Sch{\"o}del}, {Najarro}, {Muzic}, \&
  {Eckart}}]{schoedel10}
{Sch{\"o}del}, R., {Najarro}, F., {Muzic}, K., \& {Eckart}, A. 2010, \aap, 511,
  A18

\bibitem[{{Shen} {et~al.}(2010){Shen}, {Rich}, {Kormendy}, {Howard}, {De
  Propris}, \& {Kunder}}]{shen10}
{Shen}, J., {Rich}, R.~M., {Kormendy}, J., {et~al.} 2010, \apjl, 720, L72

\bibitem[{{Springel} \& {Hernquist}(2005)}]{springel05}
{Springel}, V. \& {Hernquist}, L. 2005, \apjl, 622, L9

\bibitem[{{Stanek} {et~al.}(1997){Stanek}, {Udalski}, {Szymanski}, {Kaluzny},
  {Kubiak}, {Mateo}, \& {Krzeminski}}]{stanek97}
{Stanek}, K.~Z., {Udalski}, A., {Szymanski}, M., {et~al.} 1997, \apj, 477, 163

\bibitem[{{Tacconi} {et~al.}(2010){Tacconi}, {Genzel}, {Neri}, {Cox}, {Cooper},
  {Shapiro}, {Bolatto}, {Bouch{\'e}}, {Bournaud}, {Burkert}, {Combes},
  {Comerford}, {Davis}, {Schreiber}, {Garcia-Burillo}, {Gracia-Carpio}, {Lutz},
  {Naab}, {Omont}, {Shapley}, {Sternberg}, \& {Weiner}}]{tacconi10}
{Tacconi}, L.~J., {Genzel}, R., {Neri}, R., {et~al.} 2010, 463, 781

\bibitem[{{Toomre}(1977)}]{toomre77}
{Toomre}, A. 1977, 15, 437

\bibitem[{{Udalski}(2003)}]{udalski03}
{Udalski}, A. 2003, \apj, 590, 284

\bibitem[{{Uttenthaler} {et~al.}(2012){Uttenthaler}, {Schultheis}, {Nataf},
  {Robin}, {Lebzelter}, \& {Chen}}]{uttenthaler12}
{Uttenthaler}, S., {Schultheis}, M., {Nataf}, D.~M., {et~al.} 2012, \aap, 546,
  A57

\bibitem[{{Valenti} {et~al.}(2007){Valenti}, {Ferraro}, \&
  {Origlia}}]{valenti07}
{Valenti}, E., {Ferraro}, F.~R., \& {Origlia}, L. 2007, \aj, 133, 1287

\bibitem[{{van Helshoecht} \& {Groenewegen}(2007)}]{vanhel07}
{van Helshoecht}, V. \& {Groenewegen}, M.~A.~T. 2007, \aap, 463, 559

\bibitem[{{Weinberg}(1985)}]{weinberg85}
{Weinberg}, M.~D. 1985, \mnras, 213, 451

\bibitem[{{Zoccali} {et~al.}(2008){Zoccali}, {Hill}, {Lecureur}, {Barbuy},
  {Renzini}, {Minniti}, {G{\'o}mez}, \& {Ortolani}}]{zoccali08}
{Zoccali}, M., {Hill}, V., {Lecureur}, A., {et~al.} 2008, \aap, 486, 177

\bibitem[{{Zoccali} {et~al.}(2003){Zoccali}, {Renzini}, {Ortolani}, {Greggio},
  {Saviane}, {Cassisi}, {Rejkuba}, {Barbuy}, {Rich}, \& {Bica}}]{zoccali03}
{Zoccali}, M., {Renzini}, A., {Ortolani}, S., {et~al.} 2003, \aap, 399, 931

\end{thebibliography}

\end{document}